\documentclass[prb,twocolumn,oneside,bibnotes,nofootinbib]{revtex4-2}

\usepackage{amsmath,amssymb,amsfonts,mathtools}
\usepackage{braket}

\usepackage{graphicx}
\usepackage[percent]{overpic}
\usepackage{float}
\usepackage{caption}

\usepackage{xcolor}
\usepackage{tikz}
\usepackage{adjustbox}
\usepackage{changepage}
\usepackage{soul}
\usepackage[normalem]{ulem}

\usepackage[hidelinks]{hyperref}

\begin{document}

\title{Squeezed quantum multiplets: properties and phase space representation}

\author{Juan Pablo Paz$^{1,2}$}
\author{Corina Révora$^{1,2}$}
\author{Christian Tomás Schmiegelow$^{1,2}$}

\affiliation{1. Universidad de Buenos Aires, FCEyN, Departamento de F\'{\i}sica. Buenos Aires, Argentina}

\affiliation{2. CONICET-Universidad de Buenos Aires, IFIBA, Buenos Aires, Argentina.}

\date{\today}

\begin{abstract}

We define and study the properties of ``squeezed quantum multiplets''. Ordinary multiplets are sets of $D$-orthonormal quantum states formed by superpositions of states squeezed along $D$ equally spaced directions in quadrature space. More generally, we also discuss superpositions of ``higher-order squeezed states'', including tri-squeezed and quad-squeezed states.
All these states involve superpositions of multiples of $p$ photons. We compare states in ordinary ($p=2$) multiplets and higher-order ones ($p>2$) in the most relevant cases, showing that ordinary squeezed multiplets and higher-order ones share some important similarities, as well as some differences. Finally, we present analytical expressions for phase-space distributions (Wigner and characteristic functions) representing ordinary squeezed multiplets. We use this to show that some squeezed multiplets are highly sensitive to perturbations in all phase-space directions, making them interesting for metrological applications.

\end{abstract}

\maketitle

\section{Introduction}\label{sec:introduccion}

Achieving quantum information processing is one of the most important technological challenges being pursued today~\cite{nielsen2010quantum}. On the road that may eventually lead us to building a practically useful quantum computer, different techniques are being investigated. At least two of the most important ones combine the use of individual quantum systems with a finite-dimensional space of states (qubits in the two-dimensional case and qudits in the more general one) with other components having an infinite-dimensional space of states. This is the case of an ion trap~\cite{leibfried2003quantum, haffner2008quantum}, where the internal levels of ions are used to store quantum information in a discrete manner (in the sense that they are described by a finite-dimensional space of states), and the motional degrees of freedom resemble a harmonic oscillator whose space of states is infinite-dimensional. Similarly, in the so-called circuit-QED approach~\cite{ripoll2022quantum,blais2021circuit}, the qubits (which are built using artificial atoms composed of arrays of Josephson junctions of different types) are coupled to resonators. The artificial atoms have a finite-dimensional effective space of states, while the resonators have an infinite-dimensional one. In this context, the issue of encoding, protecting, and processing quantum information in hybrid systems (which combine components having both finite- and infinite-dimensional spaces of states) arises.

The use of continuous variables to encode and process quantum information has been considered for a long time~\cite{lloyd1999quantum}. In this context, performing universal computation (or simulations) requires the creation of non-Gaussian states (which are always characterized by negativities in their Wigner function). In turn, the preparation of this kind of states requires the use of some degree of nonlinearity. This can be obtained either from nonlinear terms in the Hamiltonian or by coupling the oscillator to a qubit. In that case, one needs to perform a controlled gate, such as a controlled displacement~\cite{monroe1996schrodinger, eickbusch2022fast} or controlled squeezing~\cite{drechsler2020state,del2025controlled}, followed by measurement and post-selection. It was also recognized very early that the type of errors that continuous-variable systems can suffer is broader than in the finite-dimensional case~\cite{lloyd1999quantum}. However, powerful error-correction techniques have been developed in this context. Pioneering work by Gottesman, Kitaev, and Preskill~\cite{gottesman2000encoding} showed how to use continuous variables to encode quantum information in a robust manner by using superpositions of coherent states localized on a type of phase-space lattice. This is the well-known GKP approach, which has been the subject of extensive theoretical investigations and, most notably, has been implemented experimentally in a variety of technological setups~\cite{hu2019quantum, rosenblum2018fault, fluhmann2019encoding}.

Another important set of states of a continuous-variable system is the one defining the so-called ``rotationally invariant quantum error-correcting codes''~\cite{grimsmo2020quantum}. In this case, the set of states $\ket{\phi_m}$ used to encode quantum information forms a family with $D$ elements that are invariant under a rotation by an angle $2\pi/D$ in phase space. These states must satisfy $U(2\pi/D)\ket{\phi_m}=\exp(-i2\pi m/D)\ket{\phi_m}$, where $m=0,\ldots,(D-1)$ and $U(\theta)$ is the phase-space rotation operator by an angle $\theta$. Notably, rotational invariance implies that the states $\ket{\phi_m}$ are superpositions of Fock states containing only $m+nD$ photons, with $n$ a nonnegative integer. This property is the crucial ingredient for constructing the error-correcting code.

In this paper, we investigate a special class of rotationally invariant states. 
These states are constructed as superpositions of \(D\) quantum states that are squeezed along different directions in phase space. 
The squeezing directions are equally spaced by angles \(\theta = \pi/D\). 
We refer to these sets of states as \emph{squeezed multiplets}. 
For the simplest case (\(D = 2\)), the preparation of such states has been discussed for trapped ions~\cite{drechsler2020state} and for circuit QED~\cite{del2025controlled}. 
As we show, squeezed multiplets belong to the family of rationally invariant error-correcting codes and exhibit a well-defined behavior under parity, being even under reflections. 
Moreover, they inherit many interesting properties from their relation to squeezed states. 
Despite being highly non-Gaussian, their properties remain amenable to analytical treatment. 
In particular, we present analytic expressions for the Wigner function and the characteristic function representing these states in phase space.

We also investigate another family of rotationally invariant states using higher-order squeezed states~\cite{braunstein1987generalized}. These states are created from the vacuum by acting with an operator $S^{(p)}(r, \theta)=\exp(r(a^{p}e^{-i\theta}-a^{\dagger p} e^{i\theta}))$. This set includes the so-called trisqueezed states ($p=3$) and quadsqueezed states ($p=4$), that are invariant under a rotation by an angle $2\pi/p$ in phase space. In all cases the angle $\theta$ defines the direction in which the higher-order squeezing takes place. As opposed to ordinary squeezed multiplets, higher-order squeezed states can only be analyzed using numerical methods. It is worth noting that these states have been produced in several experiments using a variety of experimental platforms. Thus, tri- and quadsqueezed states were prepared in an ion trap~\cite{saner2024, buazuavan2024}, while trisqueezed states were investigated recently in a circuit QED experiment~\cite{chang2020observation, eriksson2024universal, agusti2020tripartite, jarvis2025observation}. In this paper, we compare the properties of the states of a squeezed multiplet and those of higher-order squeezed states, highlighting interesting differences and similarities.

The paper is organized as follows. In Section~II, the definition of squeezed multiplets is presented in detail, and some of the properties of these states are outlined. In Section~III, we consider higher-order squeezed states and introduce the concept of higher-order squeezed multiplets. In this section, we also compare the properties of ordinary squeezed multiplets and those corresponding to higher-order ones. In Section~IV, we present exact results for the Wigner and characteristic functions of any state belonging to an ordinary squeezed multiplet, showing how to obtain the analytical expressions in detail. We discuss the properties of the Wigner functions and emphasize the reasons that make these states interesting from a metrological point of view. In Section~V, we summarize our results and discuss their importance and applicability.

\section{Squeezed multiplets}\label{sec:squeez_multiplets}

%\subsection{Definitions and qudit representation}\label{sec:squeez_multiplets_A}

Here we introduce the notion of a squeezed multiplet. This is a set of $D$ mutually orthogonal states that are obtained by superposing $D$ squeezed states.

Let us start by recalling that vacuum squeezed states form a two-parameter family that is defined as $\ket{r,\theta} = S(r,\theta) \ket{0}$, where the unitary operator $S(r,\theta)$ (the squeezing operator) is
\mbox{$S(r,\theta) = \exp\left[ \frac{r}{2} \left( a^2 e^{-i\theta} - a^{\dagger 2} e^{i\theta} \right) \right]$}.
These states are eigenstates of linear combinations of creation and annihilation operators of the harmonic oscillator. In particular, $\ket{r,\theta}$ is an eigenstate, with vanishing eigenvalue, of the annihilation operator $a(r,\theta)=S(r,\theta)\,a\,S^{\dagger}(r,\theta)$, which can be expressed as
\begin{eqnarray}
a(r,\theta) 
&=&S(r,\theta)\,a\,S^{\dagger}(r,\theta)\nonumber\\
&=& a \cosh(r) + a^\dagger e^{i\theta} \sinh(r).
\end{eqnarray}

Moreover, it is relatively simple to find a way to express these states as a linear superposition of energy eigenstates. In fact, an explicit expression for the squeezed state in the number basis ($\ket{r,\theta}=\sum_n c_n\ket{n}$) can be obtained using the equation $a(r,\theta)\ket{r,\theta}=0$ and solving the recurrence relation for the coefficients. The normalized squeezed state in the number basis is then found to be:
\begin{equation}
\ket{r,\theta} = \frac{1}{\sqrt{\cosh(r)}}\sum_{k=0}^{\infty} \frac{ \sqrt{(2k)!} }{ 2^k k! } \left( -\tanh(r) \right)^k e^{i\theta k} \ket{2k}.
\label{ec:sq_state}
\end{equation}

Using superpositions of $D$ squeezed states, one can construct a set of $D$ mutually orthogonal quantum states. We denote this set as the squeezed multiplet $S_{D}$, which is formed by the states $\ket{s_{D,m}}$, with $m$ being an integer between $0$ and $D-1$, and

\begin{equation}
\ket{s_{D,m}} = N_{D,m} \sum_{j=0}^{D-1} \exp\left(-i\frac{2\pi m j}{D} \right) \ket{r, \theta_j}, 
\label{ec:sm}
\end{equation}
where $\theta_j = 2\pi j/D$ and $N_{D,m}$ is a normalization constant that can be written as

\begin{equation}
    \frac{1}{N_{D,m}^2} = \sum_{j,k=0}^{D-1}\frac{\exp\left(-i\frac{2\pi m (j-k)}{D}\right)}{\cosh(r)\sqrt{1-\tanh(r)^{2}\exp\left( i\frac{2\pi (j-k)}{D} \right)}}.
\end{equation}

That is to say, the multiplets are formed by superposing $D$ states that are squeezed along $D$ phase-space directions such that any pair of such directions forms an angle that is a multiple of $\pi/D$. The coefficients appearing in the above superposition are $D$th roots of unity and are chosen in such a way as to ensure the orthogonality between the states of the multiplet. In fact, using equation~\eqref{ec:sq_state} one can write:
\begin{equation}
\begin{split}
\ket{s_{D,m}} = \frac{N_{D,m}}{\sqrt{\cosh(r)}} &\sum_k \frac{ \sqrt{(2k)!} }{ 2^k k! } \left( -\tanh(r) \right)^k\times\\&
            \sum_{j=0}^{D-1} \exp\left( i\frac{2\pi j (k - m)}{D} \right)\ket{2k} .
\end{split}
\end{equation}

It is simple to transform this expression by performing the summation over $j$ (for any fixed value of $k$) and, in this way, show that the coefficient accompanying the state $\ket{2k}$ is non-zero if and only if $(k-m)$ is an integer multiple of $D$ (i.e.\ $k=m+nD$, with $n=0,1,\ldots$). Therefore, each member of the squeezed multiplet involves superpositions of states with different photon numbers. Thus, the state $\ket{s_{D,m}}$ is a superposition of states with photon number equal to $2nD+2m$, for any integer $n$. As a consequence, all states of the multiplet are orthogonal to each other (and their normalization is ensured by the constant $N_{D,m}$ defined above). Notice that the vacuum state is populated only for the $m=0$ state of the multiplet.

We can use the multiplet $S_D$ to represent the states of a qudit and, as each multiplet state is built from photon states that differ in number by at least two photons, the loss of a single photon can be simply detected by a parity measurement. These states can be used in a different way in order to make sure that losses of more than one photon are also detectable. In fact, for even values of $D$, we can use the $m=0$ and $m=D/2$ members of the multiplet to represent a single qubit. By doing this, the loss of up to $D-1$ photons can be detected by means of super-parity measurements. Thus, when the photon number is a multiple of $D$, we identify the absence of errors. Moreover, by detecting that the photon number is a multiple of $D-1$, we identify a single-photon loss, etc.

It is interesting to note that some of the properties of the squeezed multiplets directly follow from their behavior under phase-space rotations. As mentioned above, the rotation operator by an angle $\phi$ is $U(\phi)=\exp(-i \phi N)$, where $N=a^{\dagger} a$ (and corresponds to the time-evolution operator for a time such that $\omega t=\phi$). 
As any vacuum squeezed state is obtained by acting on the vacuum with a squeezing operator, which is the exponential of a term involving the square of the creation and annihilation operators, these states are invariant under $U(\pi)$. In turn, this implies that any squeezed state is a superposition of even-numbered eigenstates $\ket{n}$. Moreover, under a rotation by an angle $\phi=\pi/D$, the rotation operator $U(\pi/D)$ transforms any state $\ket{r, \theta_{j}}$ into the state $\ket{r, \theta_{j-1}}$.

Using this, we can show that any state in the multiplet satisfies the following identity:
\begin{equation}
    U(\pi/D) \ket{s_{D,m}}=\exp\left(-i\frac{2\pi m}{D}\right) \ket{s_{D,m}}.
    \label{eq:rot_eigenvalue}
\end{equation}
In this sense, the multiplet $S_D$ is invariant under the rotation $U(\pi/D)$.

Finally, let us mention that the behavior of the multiplet under rotations determines the number states $\ket{2k}$ that can appear in each state $\ket{s_{D,m}}$. The eigenvalue condition in Eq.~\eqref{eq:rot_eigenvalue} implies that $k$ must be of the form $k=m+nD$ for any non-negative integer $n$. This can be shown by writing $\ket{s_{D,m}}$ as a superposition of the form $\ket{s_{D,m}}=\sum_k C_k \ket{2k}$ and then imposing the eigenvalue condition, which implies that the coefficient $C_k$ must satisfy
$C_k(e^{-i2 \pi  m/D}-e^{-i 2k\pi/D})=0$.
This implies that the coefficient $C_k$ must vanish unless the equation $k=m+nD$ is satisfied, with $n=0,1,\ldots$.

\section{Higher order squeezed states and higher order squeezed multiplets}\label{sec:comparision_with_W}

In this section, we generalize the notion of squeezed multiplets to the case of higher-order squeezed states, which are defined~\cite{braunstein1987generalized} as follows: we consider the unitary operator $
%U_p(r,\theta)=\exp\left[r (a^p e^{-i\theta} - a^{\dagger p} e^{i\theta})\right]$
U_p(r,\theta)=\exp\left[r (a^p e^{-i\theta} - a^{\dagger p} e^{i\theta})\right]$. For $p=2$, it coincides with the ordinary squeezing operator. When this operator acts on the vacuum, it generates what we denote as a $p$th-order squeezed state, which is characterized by the squeezing factor $r$ and the squeezing angle $\theta$. These states will be denoted as
\begin{equation}
    \ket{r,\theta}^{(p)}=U_p(r,\theta) \ket{0}.
\end{equation}
It is clear that these states generalize the notion of squeezed states to higher order, and their creation necessarily involves some type of nonlinear interaction.

An argument analogous to that of the previous section shows that any $p$th-order squeezed state is invariant under a rotation by an angle $\phi=2\pi/p$. This symmetry arises from the fact that the $p$th-order squeezed state is generated from a unitary operator whose exponential involves the $p$th power of creation and annihilation operators. As such powers are invariant under rotations by an angle $2\pi/p$, any $p$th-order squeezed state is an eigenstate with unit eigenvalue of $U(2\pi/p)$. This, in turn, implies that a $p$th-order squeezed state can be written as a superposition of number eigenstates $\ket{kp}$, which are integer multiples of $p$.

Following the same reasoning as in the previous section, we define the $p$th-order squeezed multiplet $S_D^{(p)}$ as a set of $D$ orthogonal states. Each state in the multiplet is a superposition of $D$ $p$th-order squeezed states having a squeezing angle $\theta_j=2\pi j/D$. Thus, a $p$th-order squeezed multiplet $S^{(p)}_D$ is formed by the states:

\begin{equation}
   \ket{s_{D,m}}^{(p)}=N_{D,m}^{(p)}\sum_{j = 0}^{D-1} \exp\left(-i\frac{\, 2 \pi m j}{D}\right) \ket{r, \theta_j}^{(p)},
\end{equation}
where $N_{D,m}^{(p)}$ is a normalization constant.

As was the case for $p=2$, the states in the $p$th-order squeezed multiplet satisfy the eigenvalue equation:
\begin{equation}
U\left(\frac{2 \pi}{pD}\right)\ket{s_{D,m}}^{(p)}= \exp \left(- i\frac{2 \pi  m}{D} \right)\ket{s_{D,m}}^{(p)}.
\end{equation}

Notably, the $p$th-order squeezed multiplet can also be used to represent a qudit, since it is formed by $D$ orthonormal states. For the same reasons discussed above, any state in the multiplet is obtained as a superposition of number eigenstates $\ket{pk}$, where $k$ must satisfy $k=m+nD$ for any non-negative integer $n$. 

The higher-order squeezed multiplets have an advantage over the ones defined for $p=2$: they are separated by more than two photons (or phonons), and therefore more errors induced by photon losses are detectable by some kind of super-parity measurement. The disadvantage, on the other hand, is that they can only be studied using numerical techniques, since very few analytical results are available.

By varying the values of $p$ (the integer defining the order of the multiplet) and $D$ (the one defining its dimensionality), one can observe that there are similarities (and differences) between the ordinary squeezed multiplets (defined for $p=2$) and those corresponding to higher order.

We can analyze similarities and differences between states of ordinary and higher-order squeezed multiplets, which are worth studying. Although, to the best of our knowledge, the creation of ordinary $n$-squeezed states has not been experimentally realized, some higher-order squeezed states and their superpositions have been experimentally realized using two different platforms. Using trapped ions, trisqueezed ($p=3$) and quadsqueezed ($p=4$) states, as well as their superpositions, have been prepared~\cite{buazuavan2024,saner2024}. Moreover, using superconducting circuit-QED platforms, trisqueezed states have also been prepared~\cite{chang2020observation, eriksson2024universal, agusti2020tripartite, jarvis2025observation}.

\subsection{Comparison between ordinary and higher-order squeezed multiplets}\label{sec:comparison}

From the above discussion, it is clear that some ordinary squeezed multiplets and some higher-order ones are superpositions of states with the same number of photons. Therefore, it is natural to ask whether these states are similar or not. Here we perform this comparison, and we find that only in some cases the corresponding states are close to each other. In all cases, we compute the overlap between a given state of an ordinary squeezed multiplet (with squeezing factor $r$) and a state from a $p$th-order multiplet (with squeezing strength $r_p$).

\begin{figure}[H]
    \centering
  \begin{minipage}{0.3\textwidth}
        \begin{overpic}[width=\textwidth]{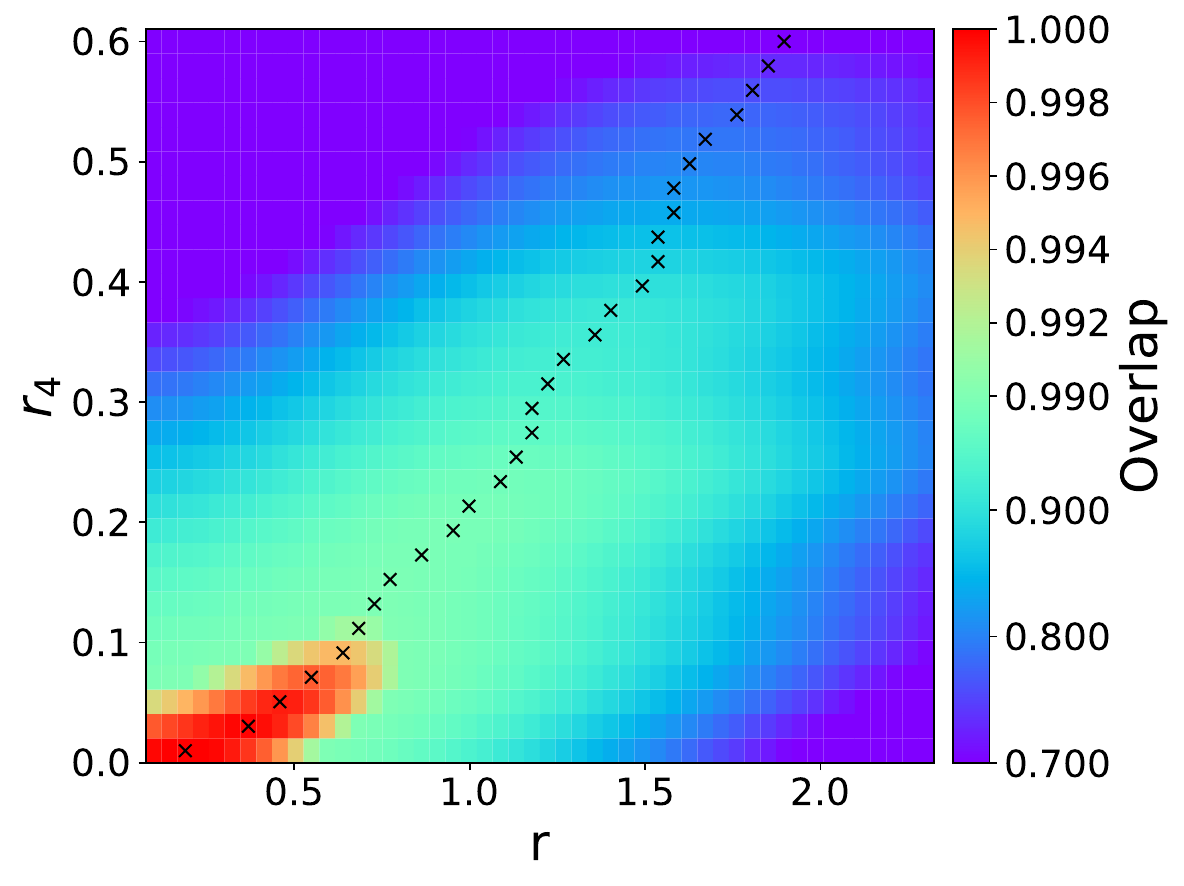}
            \put(-10,70){\textbf{(a)}}
        \end{overpic}
        \label{fig:overlap_quad2}
    \end{minipage}
    \hfill
    \begin{minipage}{0.3\textwidth}
        \begin{overpic}[width=\textwidth]{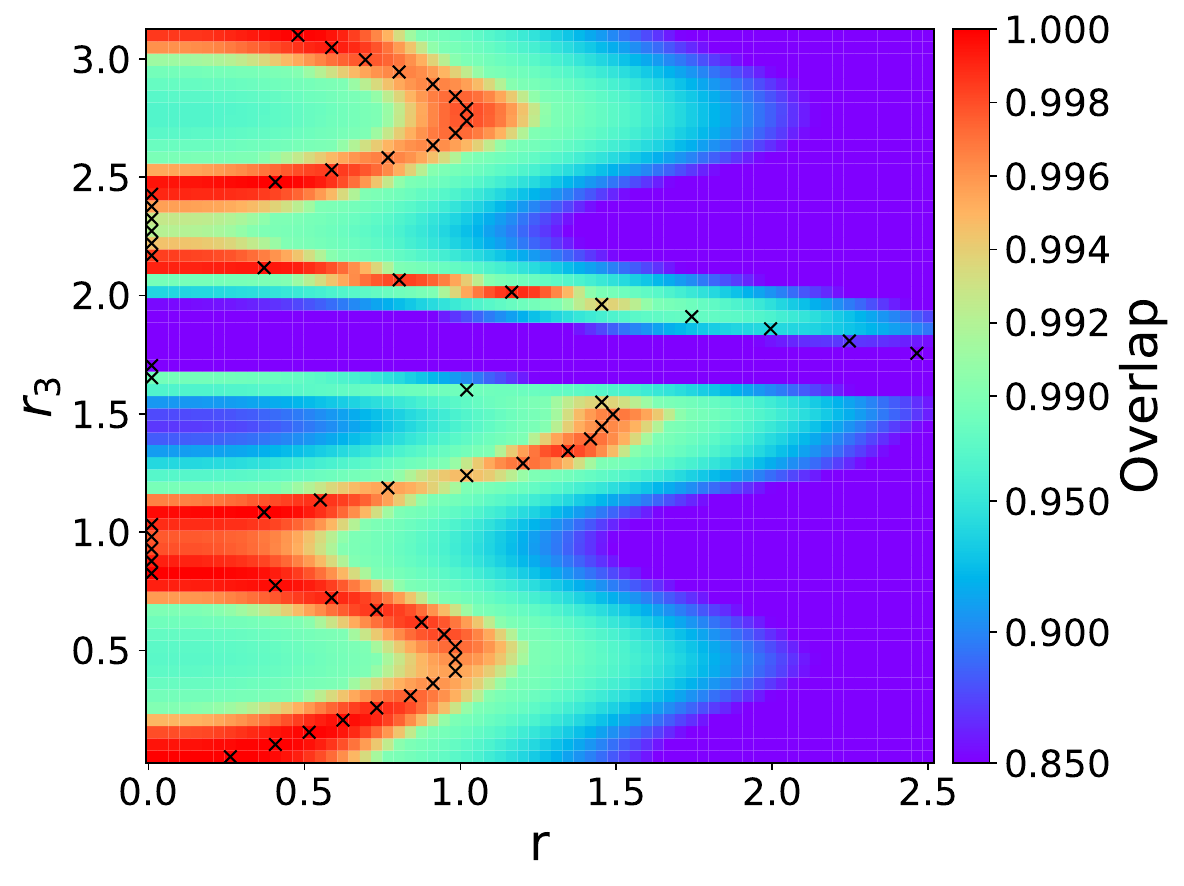}
            \put(-10,70){\textbf{(b)}}
        \end{overpic}
        \label{fig:overlap_tri}
    \end{minipage}
    \hfill
    \begin{minipage}{0.3\textwidth}
        \begin{overpic}[width=\textwidth]{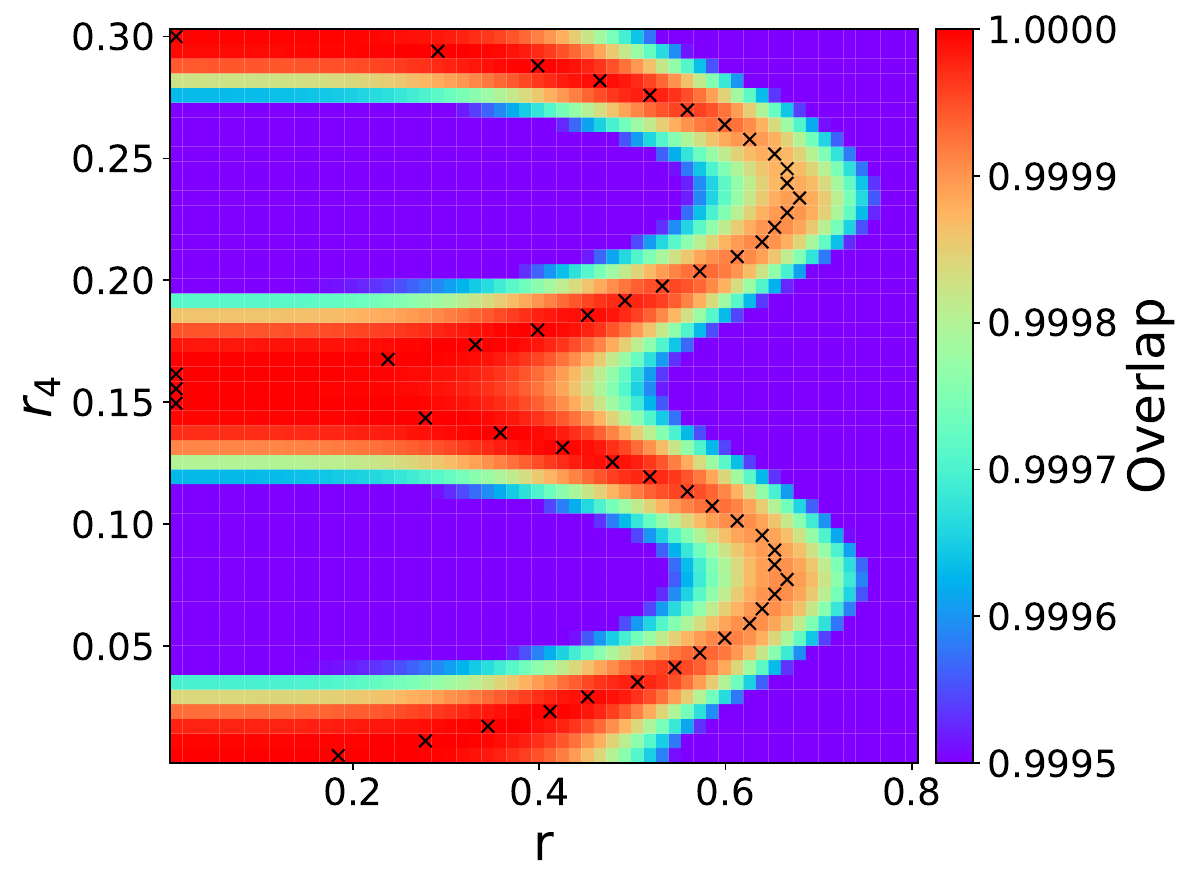}
            \put(-10,70){\textbf{(c)}}
        \end{overpic}
        \label{fig:overlap_cuad_par}
    \end{minipage}
    \hfill
    \begin{minipage}{0.3\textwidth}
        \begin{overpic}[width=\textwidth]{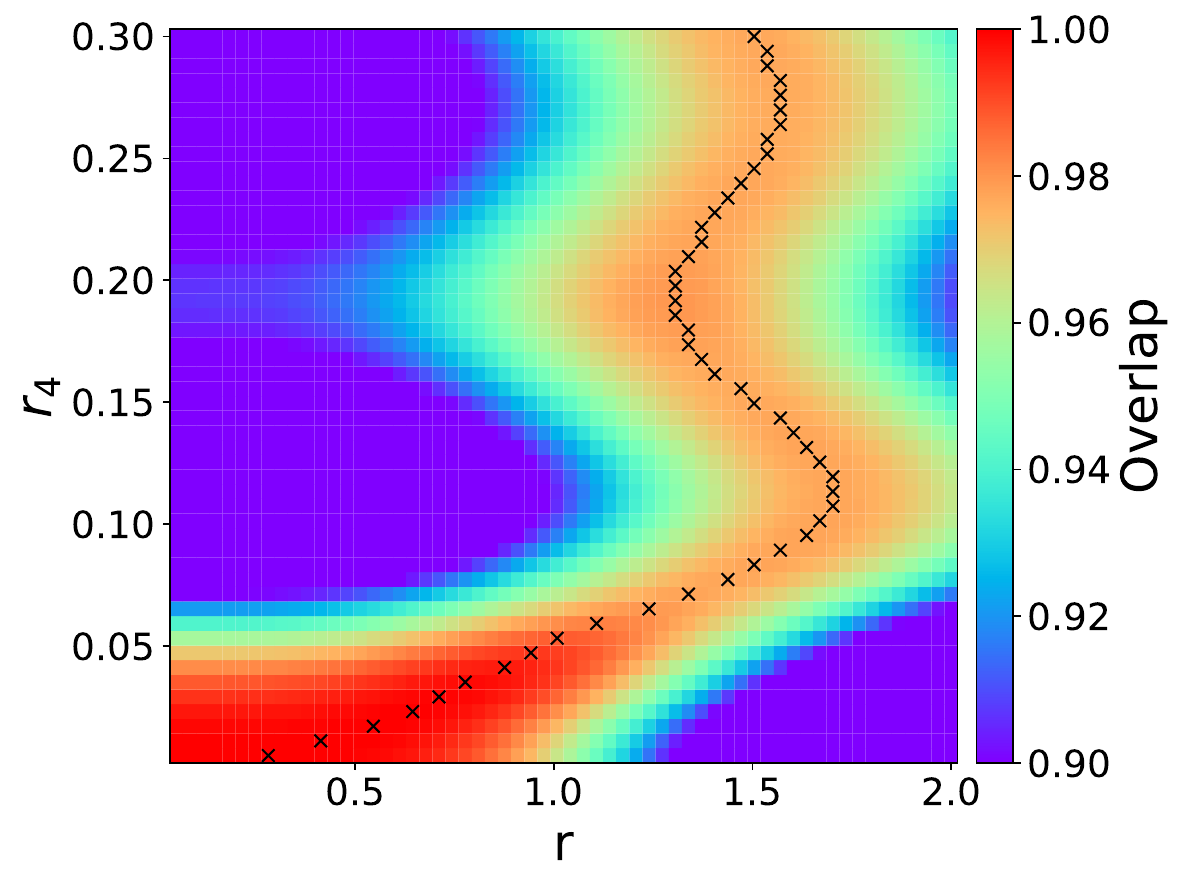}
            \put(-10,70){\textbf{(d)}}
        \end{overpic}
        \label{fig:overlap_cuad_impar}
    \end{minipage}
    
    \caption{Overlap between (a) trisqueezed states, (b) even, and (c) odd quadsqueezed states with their corresponding multiplets $D=3$, $m=0$; $D=4$, $m=0$; and $m=2$, respectively. The overlap is represented by a color scale as a function of the squeezing ($r$) and $p$th-order squeezing ($r_p$) parameters, with black crosses indicating the maximum value for each $r_p$. Note that the color scale in (a) has a different slope from $0.85$ to $0.99$ than from $0.99$ to $1$.}
    \label{fig:fig1}
\end{figure}

Let us start by analyzing the simplest case, where the states contain multiples of four photons. One such state is the even superposition of two squeezed states $\ket{s_{2,0}}$, with $m=0$ and $p=2$, $D=2$. Another example with multiples of four photons is the single quadsqueezed state $\ket{s_{1,0}}^{(4)}$, with $p=4$, $D=1$. The result of the comparison between these states is shown in Fig.~\ref{fig:fig1}(a). 
We find that the overlap is high only for small values of the squeezing strengths ($r$ and $r_4$). The overlap quickly decays below $0.9$ when $r_4$ is larger than $0.1$. The fact that the overlap is high for low values of $r$ and $r_4$ is not surprising, since in that case the states are close to the vacuum. The decay of the overlap is a clear sign that these states are different, in spite of the fact that they are superpositions of the same number states. The highest values of the overlap lie on a curve that monotonically increases as $r_4$ increases.

A different situation is observed when comparing other states. In Fig.~\ref{fig:fig1}(b), we present the overlap between states that contain multiples of six photons, $\ket{s_{3,0}}$ and $\ket{s_{2,0}}^{(3)}$. That is, the $m=0$ states of both the ordinary multiplet with $p=2$, $D=3$ and the higher-order multiplet with $p=3$, $D=2$. In this case, the overlap is larger than $0.999$ for a wider range of parameters, but displays a rather unusual behavior as a function of $r_3$. 
The highest values of the overlap do not lie on a curve that grows monotonically with $r$ and $r_3$. In turn, these curves display an irregular oscillatory behavior as a function of $r_3$. We stress that, as the higher-order states were computed numerically, special care was taken to check that this is not a numerical artifact.

The best agreement is obtained for the case of even states containing multiples of eight photons, namely $\ket{s_{4,0}}$ and $\ket{s_{2,0}}^{(4)}$. That is, for the $m=0$ states of the ordinary multiplet with $p=2$, $D=4$ and the higher-order multiplet with $p=4$, $D=2$. This is seen in Fig.~\ref{fig:fig1}(c), where it is shown that the overlap remains larger than $0.999$ for all values of $r_4$. Again, there is an irregular oscillatory behavior, whose origin still needs to be understood.

Finally, we study the case of states containing superpositions of $4+8n$ photons. These are the states $\ket{s_{4,2}}$ and $\ket{s_{2,1}}^{(4)}$, belonging to the ordinary and a quadsqueezed multiplet, respectively. The overlap is shown in Fig.~\ref{fig:fig1}(d). Here we see that the overlap values, though still high, are significantly lower than in the previous case, decaying below $0.98$.

As a conclusion, the properties of higher-order squeezed multiplets, which must be studied numerically, are rather complex. On the other hand, ordinary squeezed multiplets, which are highly non-Gaussian states, are amenable to a full analytical treatment. Both classes of states, as mentioned above, can be used to encode quantum information in an error-detectable way and, in this sense, share an important similarity. Taking this into account, we advocate for the use of ordinary squeezed multiplets in experiments, exploiting the power of exact theoretical predictions.

In the following, we show how to analytically compute the phase-space representation of these states, displaying their sensitivity to perturbations, among other properties.

\section{Phase-space representation of squeezed multiplets}\label{sec:space_rep}

Here we present a general method to analytically compute the Wigner function, $W(\alpha)=\text{Tr}(\rho D(\alpha)RD^{\dagger}(\alpha))/\pi$, and the characteristic function, $C(\alpha)=\text{Tr}(\rho D(\alpha))$, of any state within the ordinary ($p=2$) squeezed multiplet $S_D$, composed of the states $\ket{s_{D,m}}$ (see Eq.~\ref{ec:sm}). All these states are parity eigenstates, and therefore these functions are related by $W(\alpha)=C(2\alpha)/\pi$.

Rotational invariance of the squeezed multiplet imposes some important properties on $C(\alpha)$. As any state $\ket{s_{D,m}}$ is an eigenstate of the rotation operator $U(j\pi/D)$ (for any integer $j$), its characteristic function is invariant under rotations by any angle $j\pi/D$, i.e.,
\begin{equation}
    C(\alpha) = C(\alpha e^{-ij \pi/D}).
    \label{ec:rot_sym}
\end{equation}
The above equation implies that the characteristic function exhibits a periodic pattern, repeating its value for each angle $j\pi/D$. We can define a partition of phase space into slices by taking the lines forming an angle $j\pi/D$ with respect to the horizontal axis. The characteristic function is the same in any of these slices.

As parity symmetry implies that $C(\alpha)$ is real, $C(\alpha)=C(\alpha^*)$, the characteristic function must be symmetric under reflections along the horizontal axis. This property, combined with the rotational symmetry discussed above, implies that $C(\alpha)$ must also be symmetric under reflections along the bisectrix of each slice.
\subsection{An analytic expression for the characteristic function}\label{analytic_c_alpha}

Using the explicit form of $\ket{s_{D,m}}$ from Eq.~\ref{ec:sm}, its characteristic function is obtained as:
\begin{equation}
\begin{split}
C_{D,m}(\alpha) &=\\ 
N_{D,m}^2& \sum_{a,b} \exp\left({i \frac{2\pi m (a - b)}{D}}\right) \bra{r,\theta_a} D(\alpha) \ket{r,\theta_b}.
\label{ec:c_tot_sintruco}
\end{split}
\end{equation}
This expression contains $D^{2}$ terms, of which $D$ are diagonal ($a=b$) and $D(D-1)$ are non-diagonal ($a\neq b$). In the following, we compute them separately, writing $C_{D,m}(\alpha)=C_{D,m}^{(\mathrm{diag})}(\alpha)+C_{D,m}^{(\mathrm{nondiag})}(\alpha)$.

\subsubsection{Diagonal terms}

The diagonal terms can be written as: 

\begin{equation}
C_{D,m}^{(diag)}(\alpha) = N_{D,m}^2 \sum_{b} \bra{r,\theta_b} D(\alpha) \ket{r,\theta_b}.
\label{ec:c_diag}
\end{equation}
To compute this, we can first note that $\ket{r,\theta_b}=U(\theta_b/2)\ket{r,0}$. Therefore

\begin{equation}
\begin{split}
    C_{D,m}^{(diag)}(\alpha) &= N_{D,m}^2 \sum_{b} \bra{r,0}U^{\dagger}(\theta_b/2) D(\alpha) U(\theta_b/2)\ket{r,0}
    \\&=N_{D,m}^2 \sum_{b} \bra{r,0} D(\alpha e^{-i\theta_b/2})\ket{r,0}.
\end{split}
\label{ec:c_diag2}
\end{equation}

Now we can use the fact that $\ket{r,0}=S(r,0)\ket{0}$. As the squeezing operator transforms the annihilation operator as $S^\dagger(r,0)\,a\,S(r,0)=\cosh(r)\,a-\sinh(r)\,a^\dagger$, the displacement operator $D(\alpha)$ is transformed as
$S^\dagger(r,0) D(\alpha) S(r,0) = D(\alpha_r)$,
where $\alpha_r=\alpha \cosh(r)+\alpha^* \sinh(r)$. Using this, the diagonal terms can be written as:
\begin{equation}
\begin{split}
    C_{D,m}^{(diag)}(\alpha) &=N_{D,m}^2 \sum_{b} \bra{0} D(\alpha_{r, \theta_b})\ket{0},
\end{split}
\label{ec:c_diag3}
\end{equation}
where $\alpha_{r,\theta_b} = \alpha e^{-i\theta_b/2} \cosh(r) + \alpha^* e^{i\theta_b/2} \sinh(r)$. Now, writing the complex amplitude as a function of the phase-space coordinates as $\alpha=x+ip$, we have: 
\begin{equation}
    \alpha_{r,\theta_b} = e^{r}x_b + i e^{-r} p_b, 
    \label{eq:alpharxp}
\end{equation}
where $x_b = x \cos(\theta_b/2) + p \sin(\theta_b/2)$ and $p_b = -x \sin(\theta_b/2) + p \cos(\theta_b/2)$ are the position and momentum coordinates with respect to an axis rotated by an angle $\theta_b/2$.

Next, using the fact that $D(\alpha_{r,\theta_b})\ket{0}=\ket{\alpha_{r,\theta_b}}$ is the coherent state with complex amplitude $\alpha_{r,\theta_b}$, and that the overlap between this state and the vacuum is $\braket{0|\alpha_{r,\theta_b}}=\exp\{-|\alpha_{r,\theta_b}|^2/2\}$, the diagonal term can be written as:
\begin{equation}
\begin{split}
    C_{D,m}^{(diag)}(\alpha) &=N_{D,m}^2 \sum_{b} \exp\left(-|\alpha_{r,\theta_b}|^2/2\right)  
    \\&=N_{D,m}^2 \sum_{b} \exp\left(-\frac{x_b^2 e^{2r}+p_b e^{-2r}}{2}\right).
\end{split}
\label{ec:c_diag_final}
\end{equation}

We see that the diagonal term is composed of a sum of exponentials that are constant along different ellipses in phase space. In fact, the contours of constant $|\alpha_{r,\theta_b}|^2$ correspond to ellipses rotated by an angle $\theta_b/2$ in phase space, whose main axes are contracted along the $x_b$ direction and elongated along the $p_b$ direction.

\subsubsection{Non-diagonal terms}

We now compute the $D(D-1)$ non-diagonal terms. To do so, we note that the behavior of $\ket{r,\theta_b}$ under parity can be used to show that
$\bra{r,\theta_a} D(\alpha) \ket{r,\theta_b}=\bra{r,\theta_b} D(\alpha) \ket{r,\theta_a}^{*}$.
This implies that the sum in the non-diagonal term can be restricted to the cases where $b\leq a$, obtaining:

\begin{equation}
\begin{split}
C_{D,m}^{(nondiag)}(\alpha) 
=& 2N_{D,m}^2 \times\\
\sum_{a}^{D-1} \sum_{b<a} \text{Re}&\left[\exp\left(\frac{2\pi m (a - b)}{D}\right)\bra{r,\theta_a} D(\alpha) \ket{r,\theta_b}\right].
\end{split}
\label{ec:c_nondiag_1er_truco}
\end{equation}

Following the same strategy, we again use the facts that $\ket{r,\theta_b}=U(\theta_b/2)\ket{r,0}$ and $\ket{r,0}=S(r,0)\ket{0}$, as well as the properties of the displacement operator under squeezing described above, to write the overlap term as:
\begin{equation}
\begin{split}
    \bra{r,\theta_a} &D(\alpha) \ket{r,\theta_b} =\\
    =&\bra{r,\theta_{a}-\theta_{b}} D(\alpha e^{-i\theta_b/2}) \ket{r,0}\\
    =&\bra{r,\theta_{a}-\theta_{b}}S(r,0)S^\dagger(r,0) D(\alpha e^{-i\theta_b/2}) S(r,0)\ket{0}\\
    =&\bra{r,\theta_{a}-\theta_{b}}S(r,0) D(\alpha_{r,\theta_b} )\ket{0}\\
     =&\bra{r,\theta_{a}-\theta_{b}} S(r,0) \ket{\alpha_{r\theta_b}}\\
     = &\braket{\phi(r,\theta_{ab})|\alpha_{r\theta_b}},
\end{split}
\end{equation}
where, in the last step, we defined the state $\ket{\phi(r,\theta_{ab})} = S^{\dagger}(r,0)\ket{r,\theta_{a}-\theta_b}$. Therefore, to obtain the off-diagonal terms, we need to compute the overlap between the coherent state $\ket{\alpha_{r,\theta_b}}$ and the state $\ket{\phi(r,\theta_{ab})}$. This second state is \textit{not} a simple squeezed state, as it is the composition of two squeezing operators that together do \textit{not} form a squeezing operator.

To obtain $\ket{\phi(r,\theta_{ab})}$, we can use the fact that the state $\ket{r,\theta_{ab}}$ is an eigenstate with zero eigenvalue of the operator $a(r,\theta_{ab})$, defined as
$a(r,\theta_{ab})= \cosh(r) a + \sinh(r) e^{i \theta_{ab}} a^\dagger$. Using this, we can show that the state $\ket{\phi(r,\theta_{ab})}$ is itself an eigenstate with zero eigenvalue of the operator $\tilde{a}(r,\theta_{ab})=S^\dagger(r,0)a(r,\theta_{ab})S(r,0)$, where

\begin{equation}
\begin{split}
    \tilde{a}(r,\theta_{ab}) &= \cosh(r) a(r,\theta_{ab}) - \sinh(r) a^\dagger(r,\theta_{ab}) \\&= u(r,\theta_{ab}) a + v(r,\theta_{ab}) a^\dagger.
\end{split}
\end{equation}
Here, the coefficients $u(r,\theta_{ab})$ and $v(r,\theta_{ab})$ are such that
\begin{equation}
\begin{split}
    &u(r,\theta_{ab})= e^{-i\theta_{ab}/2}\left(\cos\left(\frac{\theta_{ab}}{2}\right)+i\sin\left(\frac{\theta_{ab}}{2}\right) \cosh(2r)\right), \quad \\&v(r,\theta_{ab})= i e^{i \theta_{ab}/2} \sin\left(\frac{\theta_{ab}}{2}\right) \sinh(2r).
\end{split}
\end{equation}

Now, it is straightforward to obtain an expression for $\ket{\phi(r,\theta_{ab})}$ by solving the recurrence relation that is implicit in the equation $\tilde{a}(r,\theta_{ab}) \ket{\phi(r,\theta_{ab})}=0$. In this way, we find
\begin{equation}
\ket{\phi(r,\theta_{ab})} = K_{\theta ab} \sum_{k=0}^\infty \frac{ \sqrt{(2k)!} }{ 2^k k! } \left( -\frac{v(r, \theta_{ab})}{u(r,\theta_{ab})} \right)^{k}  \ket{2k},
\end{equation}
where the normalization constant is 
\begin{equation}
\begin{split}
    K_{\theta ab} &= \left<r,\theta_a-\theta_b|r,0\right>
    \\&= \frac{1}{\cosh(r)\sqrt{1-\tanh^2(r)e^{-i(\theta_{a}-\theta_{b})}}}.
\end{split}
\end{equation}

The computation of the overlap is straightforward, as we now have the states expressed in the number basis. By rearranging terms, we obtain:
\begin{equation}
\begin{split}
    \langle{\phi(r,\theta_{ab})}\ket{\alpha_{r\theta_b}}&=\\K_{\theta ab} e^{-\frac{|\alpha_{r\theta_b}|^2}{2} } 
    &\sum_{k=0}^\infty \frac{1}{k!} \left( -\left(\frac{v(r,\theta_{ab})}{u(r, \theta_{ab})}\right)^{*}\frac{\alpha_{r\theta_b}^2 }{2} \right)^k,
\end{split}
\end{equation}
which can be rewritten by summing the Taylor series of the exponential to obtain the desired result, remembering that
$\bra{r,\theta_a} D(\alpha) \ket{r,\theta_b}=\langle{\phi(r,\theta_{ab})}\ket{\alpha_{r\theta_b}}$, as:
\begin{equation}
\begin{split}
     \bra{r,\theta_a} D(\alpha) &\ket{r,\theta_b}=\\
    K_{\theta ab} &\exp\left(
    -\frac{|\alpha_{r\theta_b}|^2}{2} 
    -\left(\frac{v(r,\theta_{ab})}{u(r, \theta_{ab})}\right)^{*}\frac{\alpha_{r\theta_b}^2 }{2} \right).
\end{split}
\label{eq:overlap_final}
\end{equation}

This is the final result from which each non-diagonal term in Eq.~\ref{ec:c_nondiag_1er_truco} can be computed. It is worth noting that, in all cases, the off-diagonal terms are the product of an exponential and a trigonometric function, which both depend on quadratic forms of the phase-space coordinates $x$ and $p$. The dependence on these coordinates enters through $\alpha_{r,\theta_b}$, as indicated in Eq.~\eqref{eq:alpharxp}. 

After some algebraic manipulations (see Appendix~\ref{app:asymptotic_limits}), we can show that each of these overlap terms oscillates along regions that are bounded by hyperbolas whose asymptotes coincide with the main axes of the ellipses associated with the states $\ket{r,\theta_a}$ and $\ket{r,\theta_b}$. Examples of the behavior of these terms will be displayed below.

\begin{figure}[b]
    \centering

    \begin{minipage}{0.4\textwidth}
        \begin{overpic}[width=\textwidth]{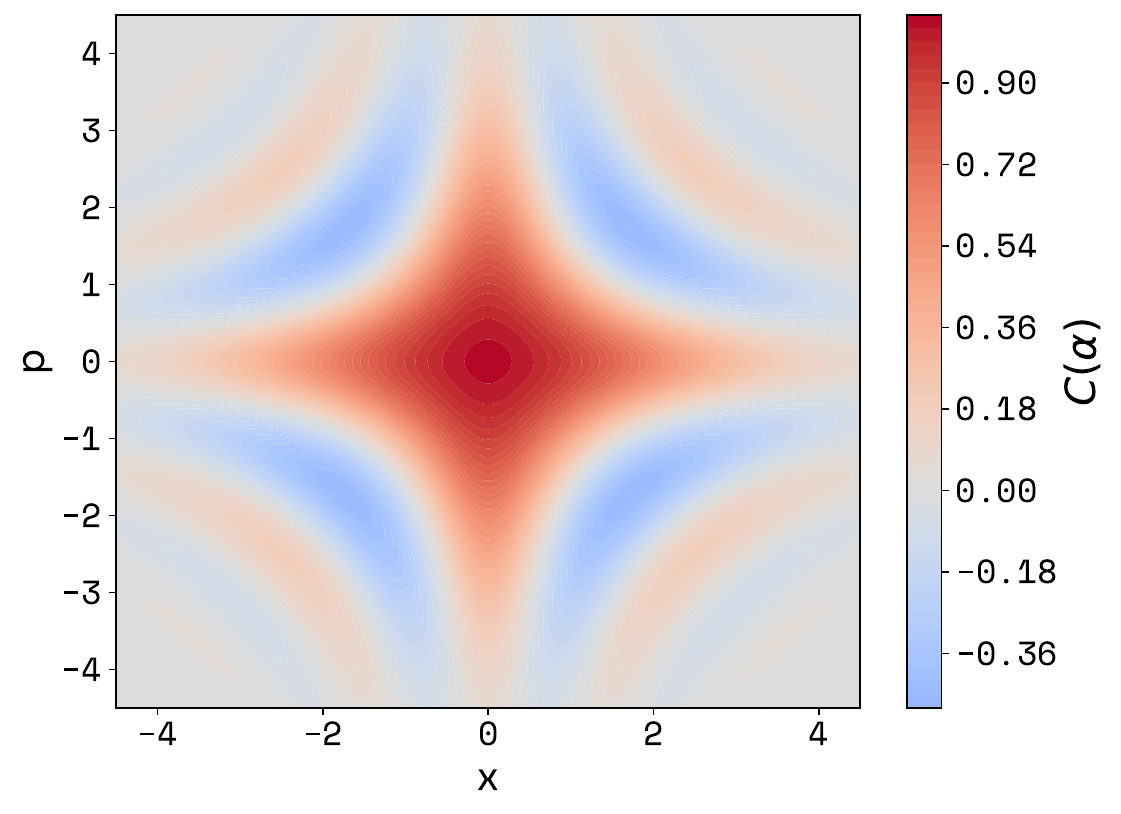}
            \put(-10,70){\textbf{(a)}}
        \end{overpic}
        
    \end{minipage}
    \hfill
    \centering

    \begin{minipage}{0.4\textwidth}
        \begin{overpic}[width=\textwidth]{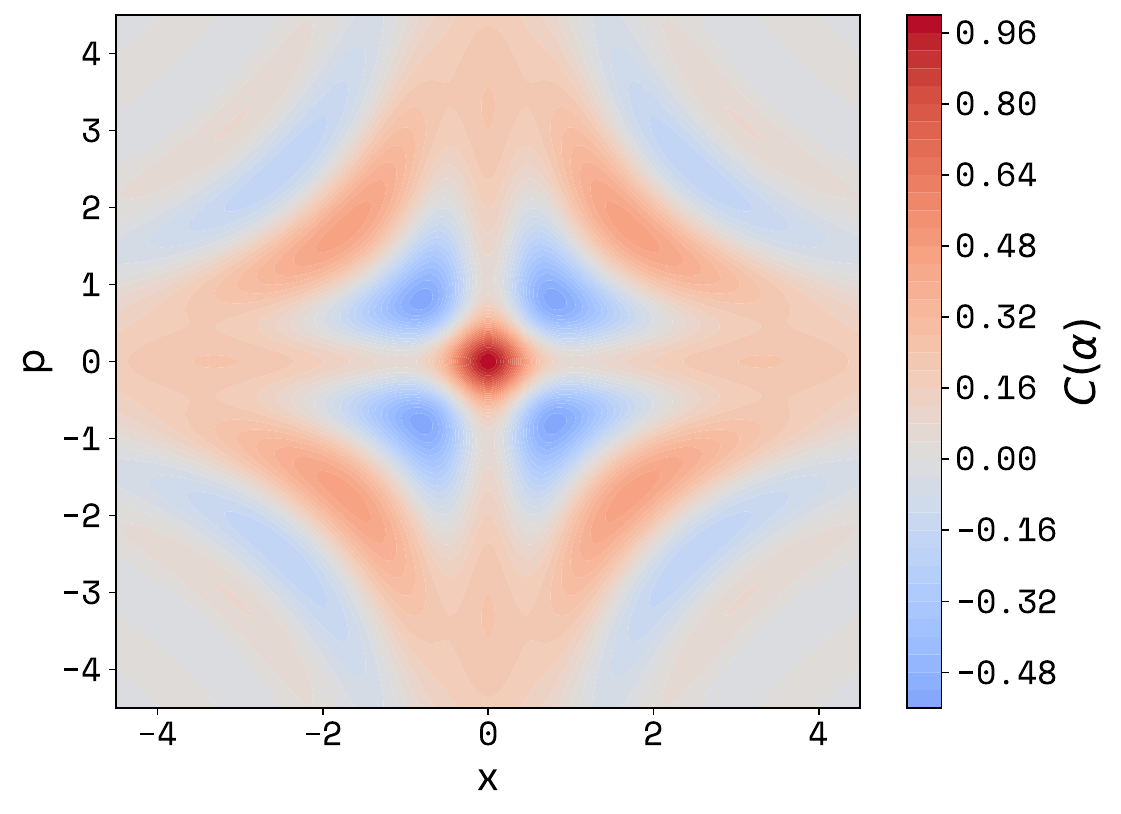}
            \put(-10,70){\textbf{(b)}}
        \end{overpic}
        
    \end{minipage}
     
    \caption{Characteristic function for $D=2$: (a) $m=0$ and (b) $m=1$. The squeezing parameter in both panels is $r=1$.}
    \label{fig:C20}
\end{figure}

\subsection{Evaluation of the characteristic functions}

To understand the behavior of $C_{D,m}(\alpha)$, it is useful to evaluate it for different values of $D$.

\subsubsection{Example 1: $D = 2$}

We start with the simplest case, $D=2$, $m=0,1$, which has been analyzed before by Drechsler et al.~\cite{drechsler2020state}. This particular case can be obtained from the expressions above by setting $\theta_{ab}=\pi$. By doing this, we obtain the characteristic function:
\begin{equation}
\begin{split}
C_{2,m}&(x,p)=\\
=&N_{2,m}^2	\Bigg(
\exp\left(-(x e^r+pe^{-r})\right)+\\
&+\exp\left(-(x e^{-r}+pe^r)\right)+ \\
	&+(-1)^m\,{{\exp\left({-{{(x^2+p^2)}\over{2\cosh(2r)}}}\right)}
		\frac{\cos\left(xp\tanh(2r)\right)}{{\sqrt{\cosh(2r)}}} }\Bigg),
\end{split}
\end{equation}
where $N_{2,m}^2=1/\left(1+(-1)^m/\sqrt{\cosh(2r)}\right)$. 

From this formula, we see that the characteristic function has a diagonal term that is the sum of two exponentials whose contours are ellipses contracted and expanded along the $x$ and $p$ axes. The interference term displays an oscillatory pattern whose values remain constant, up to a Gaussian modulation, along hyperbolas whose asymptotes are the $x$ and $p$ axes. These features can be seen in Fig.~\ref{fig:C20}, where the characteristic function $C_{2,m}(x,p)$ is shown for $r=1$.

It is worth noting a property of this example that is not valid for higher values of $D$: the hyperbolas arising from the interference term never intersect the ellipses that appear in the diagonal part $C^{(\mathrm{diag})}(\alpha)$.

\subsubsection{Example 2: $D=3$, $m=0$}

\begin{figure*}[t]  \includegraphics[width=0.95\textwidth]{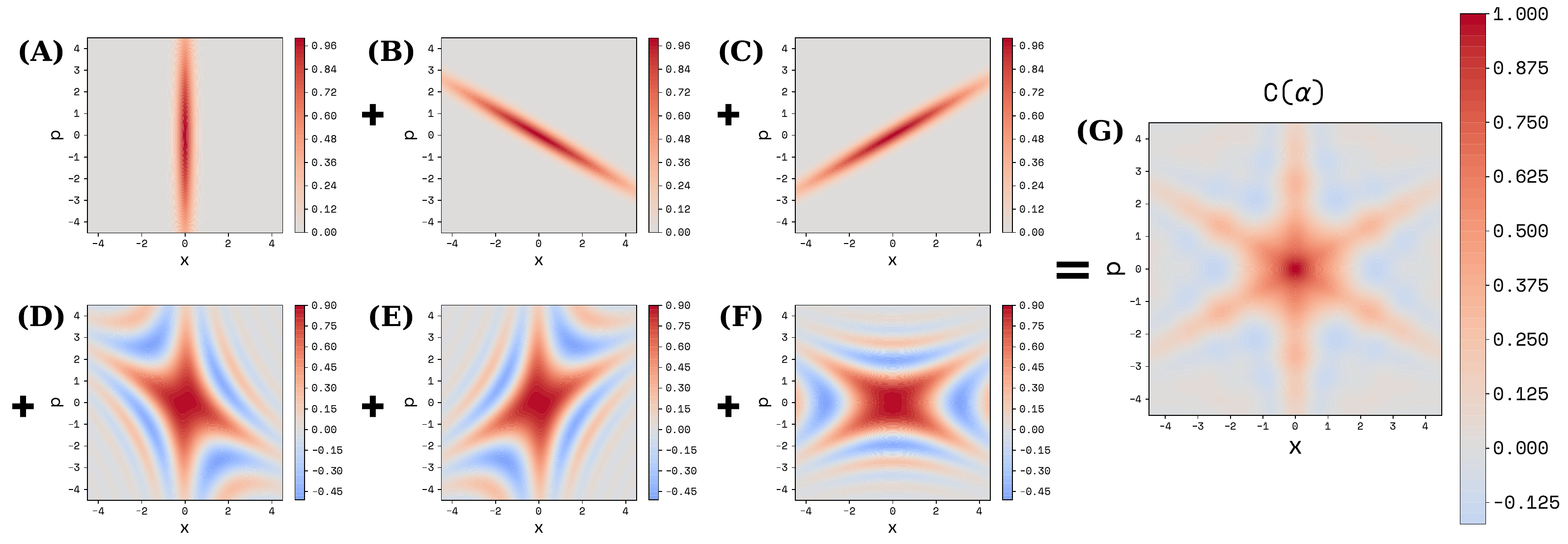}
\caption{Terms corresponding to the characteristic function of the squeezed multiplet for $D=3$, $m=0$, with $r=1.2$, shown in the panels discussed in the text. Panels (a), (b), and (c) show the direct (diagonal) terms, while panels (d), (e), and (f) correspond to the interference (non-diagonal) terms of the characteristic function. The full characteristic function, constructed as the sum of panels (a)–(f), is shown in panel (g).}
\label{fig:example_D3_C}
\end{figure*}

In this case, both the diagonal contribution $C_{3,0}^{(\mathrm{diag})}(\alpha)$ and the off-diagonal contribution $C_{3,0}^{(\mathrm{nondiag})}(\alpha)$ are given by the sum of three terms. The diagonal contribution is the sum of three exponentials that remain constant along ellipses, with principal axes oriented at angles $0$, $2\pi/3$, and $4\pi/3$, as shown in Fig.~\ref{fig:example_D3_C}(a), (b), and (c). On the other hand, the three non-diagonal terms (Eq.~\eqref{ec:c_nondiag_1er_truco}) arise from the interference between the three distinct pairs of states of the form $\ket{r,\theta_a}$ and $\ket{r,\theta_b}$, with $a\neq b$. As discussed above, these terms are rotated versions of one another, as illustrated in Fig.~\ref{fig:example_D3_C}(d), (e), and (f).

It is noticeable that the oscillatory part associated with the interference between two states intersects the elliptical region associated with the diagonal contribution of the remaining state. As mentioned above, this does not happen for the $D=2$ case. Due to this fact, the value of the total characteristic function is modulated along the three main ellipses, as seen in Fig.~\ref{fig:example_D3_C}(g). This modulation is more noticeable for higher values of $m$.

In this figure, the rotation and reflection symmetries discussed above for the general case are clearly visible.

\begin{figure*}[t]  \includegraphics[width=0.95\textwidth]{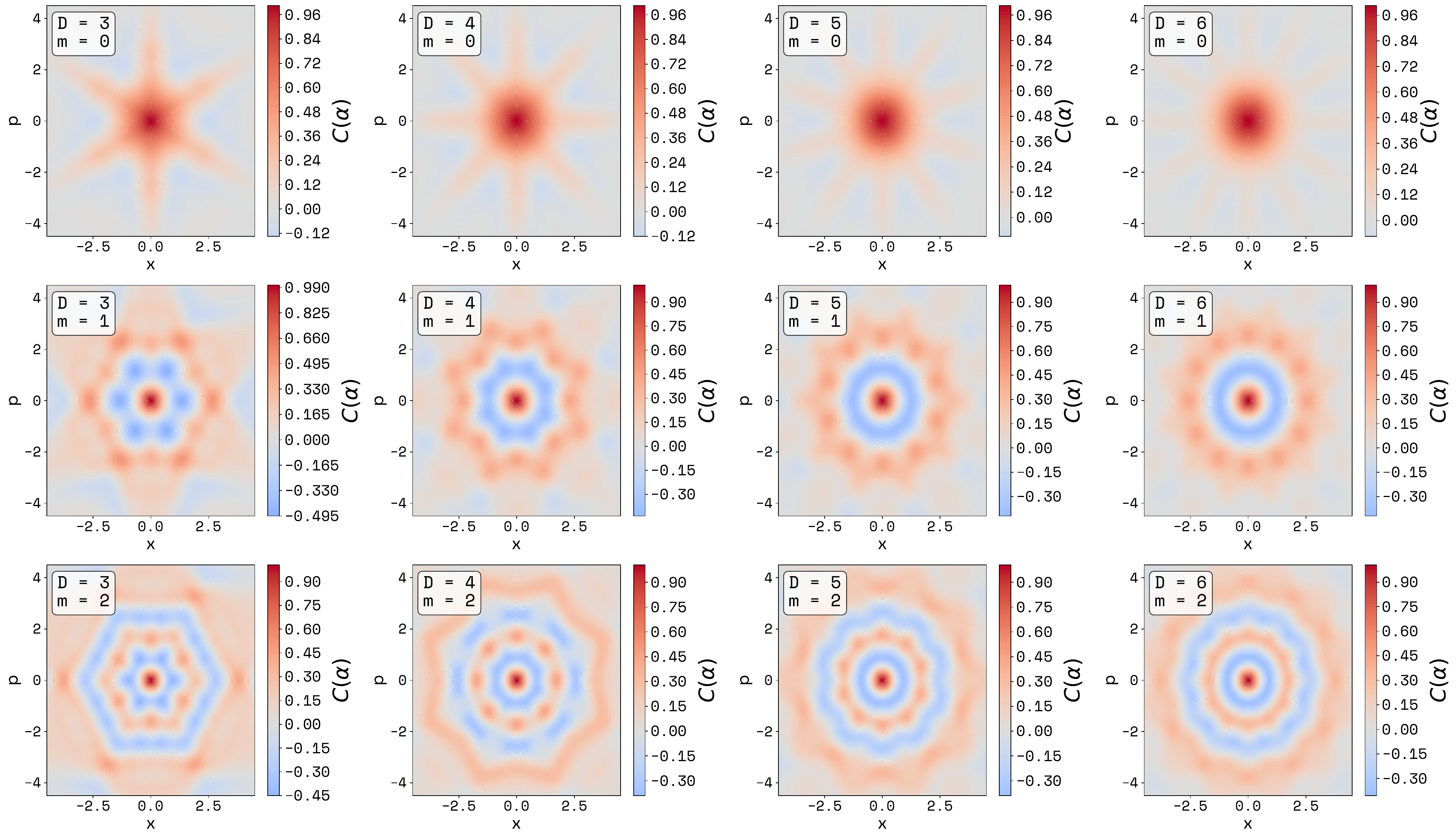}
\caption{Examples characteristic functions of various multiplets with $r=1$. Columns correspond to $D=3$, $4$, $5$, and $6$, while rows correspond to $m=0$, $1$, and $2$.}
\label{fig:example_C_multiplets}
\end{figure*}

\subsubsection{Further examples varying $D$, $m$ and $r$}

A collection of characteristic functions for different multiplet states is shown in Fig.~\ref{fig:example_C_multiplets} for $r=1$. The columns correspond to the multiplet dimensions $D=3$, $4$, $5$, and $6$, and the rows to the values of $m=0$, $1$, and $2$. In all cases, the rotation and reflection symmetries are clearly displayed. Moreover, the modulation induced by the intersection between the diagonal and non-diagonal terms is more pronounced. For the case $m\geq 1$, this modulation gives rise to a line of zeros for $C(\alpha)$ that encircles the phase-space origin. This is an important effect, since it means that the state $\ket{s_{D,m}}$ with $m\geq1$ is transformed into an orthogonal state for small displacements along any phase-space direction. This hypersensitivity to small displacements makes these states interesting for metrological applications.

\begin{figure}[h]  \includegraphics[width=0.47\textwidth]{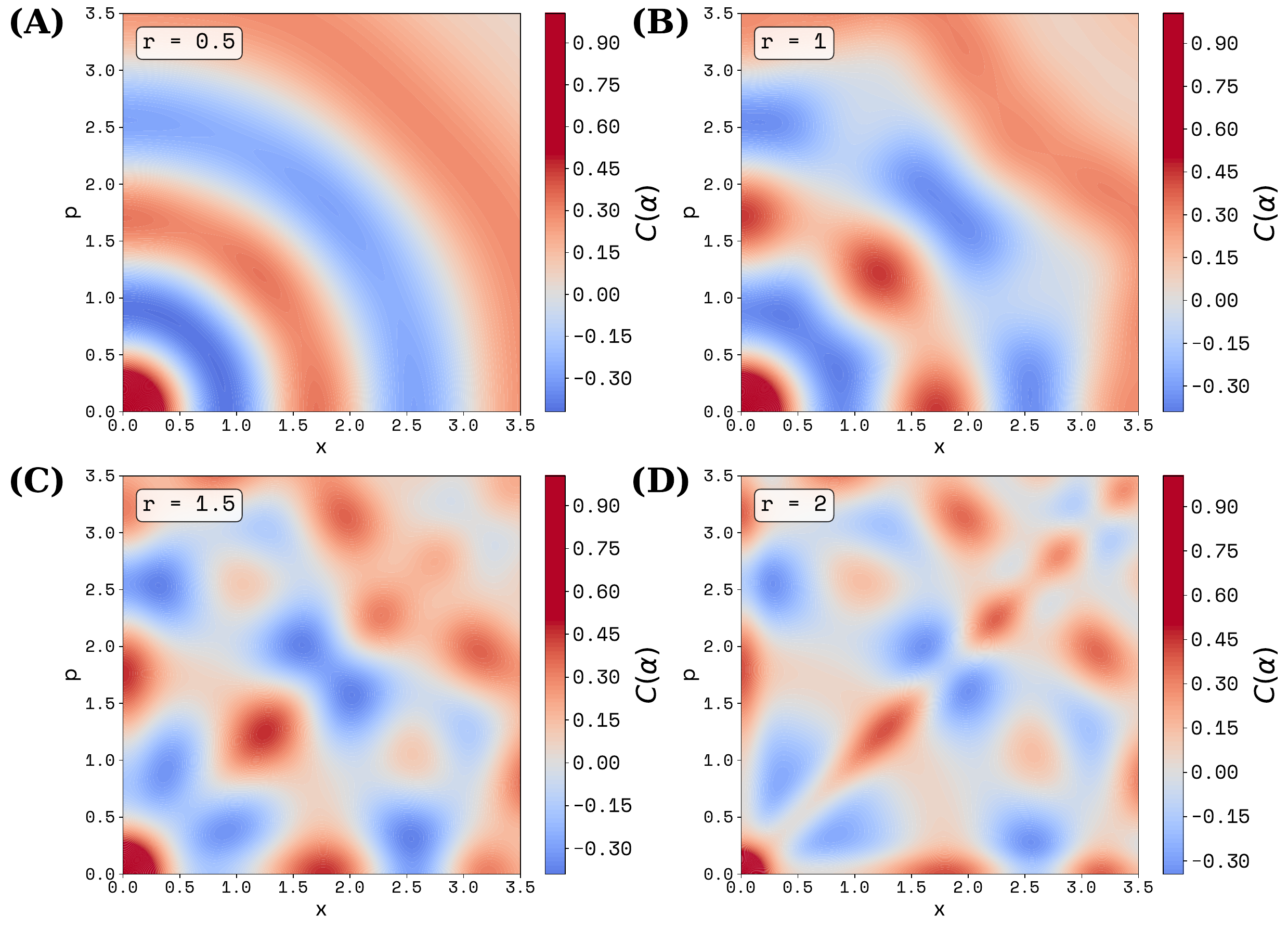}
\caption{Upper-right quadrants of the characteristic functions for increasing $r=0.5$, $1$, $1.5$, and $2$, for fixed $D=4$ and $m=2$.
}
\label{fig:example_C_rs}
\end{figure}

The squeezing strength $r$ also affects the positions of the zeros of the characteristic function. To illustrate this, in Fig.~\ref{fig:example_C_rs} we show a section of the characteristic function for $D=4$ and $m=2$, for increasing values of $r=0.5$, $1$, $1.5$, and $2$. One can see that increasing $r$ shifts the first zero of the characteristic function towards the origin. However, this comes at the cost of eliminating the continuous line of zeros surrounding the origin.

\section{Conclusions}
In this paper, we have introduced the notion of a squeezed quantum multiplet: a set $S_D$ of $D$ orthonormal states, each of which is a superposition of $D$ states squeezed along different phase-space directions. When all consecutive directions are separated by the same angle, the multiplet is rotationally invariant and is formed by eigenstates of a phase-space rotation operator with eigenvalues $\exp(-i2\pi m/D)$ (for an integer $m$). We study squeezed multiplets built using ordinary squeezed states and also higher-order multiplets. In the ordinary case, the states are linear combinations of Fock states with an even number of photons. Instead, in the higher-order case, the states involve linear combinations of Fock states with photon numbers that are multiples of $p$. As a consequence of rotational invariance, multiplet states involve linear combinations of number states with $p(m+nD)$ photons (for any integer $n$). Therefore, there are states in ordinary multiplets that are linear combinations of the same Fock states as states in higher-order multiplets. We compare these states and find that, although there are values of the squeezing strength for which the ordinary and higher-order multiplet states have a very high overlap, there are others for which the overlap is low. The properties of higher-order states, which can only be studied numerically, deserve further study.

We took advantage of the simple structure of ordinary squeezed multiplets, which allowed us to derive an analytical formula for the Wigner function of any state belonging to an ordinary squeezed multiplet. The fact that an exact result can be obtained for ordinary multiplets is an advantage over higher-order ones, which can only be studied numerically. Ordinary multiplets share the same properties as higher-order ones regarding their capability to store quantum information in a way that makes photon loss detectable by super-parity measurements. In this sense, the use of ordinary squeezed multiplets has the same benefits as higher-order ones, while also bringing the advantage of an analytical description that allows them to be studied in detail.

For example, properties that depend on the detailed structure of the phase-space distribution, such as the existence of zeros near the origin, can be more rigorously characterized. In fact, we use this to show that some ordinary multiplet states may turn out to be useful for metrology, as they display hypersensitivity to perturbations along any phase-space direction.

Finally, it is interesting to note that experimental procedures have been proposed to prepare ordinary multiplet states for the simplest ($D=2$) case in a systematic way~\cite{drechsler2020state}. For this, we require the use of a qubit that is coupled to the oscillator via a ``controlled-squeezing'' gate, followed by a projective measurement on the qubit and a post-selection process. This protocol can be extended to prepare multiplets with arbitrary values of $D$. For this, it is necessary to have a qutrit, a controlled-squeezing operation that acts only if the system is in one of the qutrit states, and a projective measurement followed by post-selection. This protocol will be studied in detail elsewhere.

\bibliographystyle{apsrev4-1} % O apsrev4-2 si lo tenés
\bibliography{biblio}

\appendix

\section{Asymptotic limit}
\label{app:asymptotic_limits}

We are interested in studying the asymptotic limit of the overlap, Eq.~\eqref{eq:overlap_final} of the main text, for different values of $D$:
\begin{equation}
\begin{split}
     \bra{r,\theta_a} D(\alpha) &\ket{r,\theta_b}=\\
    K_{\theta ab} &\exp\left(
    -\frac{|\alpha_{r\theta_b}|^2}{2} 
    -\left(\frac{v(r,\theta_{ab})}{u(r, \theta_{ab})}\right)^{*}\frac{\alpha_{r\theta_b}^2 }{2} \right).
\end{split}
\end{equation}
Clearly, it is enough to analyze the behavior of the term that appears in the exponent of this expression. In particular, we are interested in the second term, as it is the one responsible for the oscillatory behavior of the characteristic function. As argued in the text, the characteristic function repeats itself in each slice, so it is sufficient to evaluate this term for $\theta_a=2\pi/D$ and $\theta_b=0$.

For $D=2$, we have $\theta_a=\pi$. The interference term, as shown in the main text, oscillates as $\cos\!\left[\tanh(2r)\,x p\right]$. In this case, the characteristic function has zeros on the lines defined by the equation $\tanh(2r)\,x p=\frac{\pi}{2}(2n+1)$, with $n\in\mathbb{Z}$. These lines define hyperbolas that asymptotically approach the horizontal and vertical axes, which correspond to the squeezing directions.

For arbitrary values of $D$, we analyze the limit of large squeezing, i.e., $r\rightarrow\infty$. In this case, we obtain the following expression for the second term in the exponential of the overlap. Expressing it in terms of the quadratures $x$ and $p$ (using $\alpha=x+ip$), we obtain:
\begin{equation}
\begin{split}
   \left(\frac{v(r,\theta_{a})}{u(r,\theta_a)}\right)^{*}&\frac{\alpha_{r0}^2 }{2} =\\
   =&\frac{\sinh(2r) \sin(\theta_a/2)}{1+\sin^2(\theta_a/2) \sinh^2(2r)} 
   \times\\&\left(i\cos(\theta_a/2)+\sin(\theta_a/2) \cosh(2r)\right)
   \times\\&\frac{(x^2e^{2r}-p^2e^{-2r}+2ixp)}{2}.
    \label{ec:numb_argument}
\end{split}
\end{equation}
In the limit of strong squeezing $(r\rightarrow\infty)$ we have:
\begin{equation}
\begin{split}
     \left(\frac{v(r,\theta_{a})}{u(r,\theta_a)}\right)^{*}&\frac{\alpha_{r0}^2 }{2}
     \overset{\huge {r\rightarrow\infty}}{\longrightarrow}\\
      &\frac{x^2 e^{2r}}{2}
      + i \frac{x e^{2r}}{2\sin(\theta_a/2)}
      \left[\cos(\theta_a/2) x + \sin(\theta_a/2) p\right].
\label{ec:lim_r_infty}
\end{split}
\end{equation}
The imaginary part gives rise to the oscillatory behavior. From this expression we can see that the number of oscillations increases for large values of $D$, as $\theta_a=2\pi/D$. Moreover, from this equation we see that the zeros of this function lie on hyperbolas whose asymptotes are given by the equations $x=0$ and $x=-\tan(\theta_a/2)\,p$.

\end{document}